\documentclass[aps,prl,preprint,superscriptaddress,showpacs,showkeys,]{revtex4}

\usepackage{graphicx}

\usepackage{dcolumn}

\usepackage{bm}

\usepackage{amsfonts}

\usepackage{amsmath}

\usepackage{amssymb}

\usepackage{latexsym}

\usepackage{textcomp}

\usepackage{color}

\begin{document}






\preprint{spallation}
\preprint{version7}


\title{Selective Delamination upon Femtosecond Laser Ablation of Ceramic Surfaces }

\author{Frederik Kiel}

\email[Corresponding author: ]{frederik.kiel@ruhr-uni-bochum.de}

\affiliation{Applied Laser Technologies, Ruhr-Universit\"{a}t Bochum, Universit\"{a}tsstra\ss{}e 150, 44801 Bochum, Germany}

\author{Nadezhda M. Bulgakova}
\affiliation{HiLASE Centre, Institute of Physics of the Czech Academy of Sciences, Za Radnic\'i 828, 25241 Doln\'i B\v{r}e\v{z}any, Czech Republic}


\author{ Andreas Ostendorf}

\affiliation{Applied Laser Technologies, Ruhr-Universit\"{a}t Bochum, Universit\"{a}tsstra\ss{}e 150, 44801 Bochum, Germany}

\author{Evgeny L. Gurevich}

\email[Corresponding author: ]{gurevich@lat.rub.de}

\affiliation{Applied Laser Technologies, Ruhr-Universit\"{a}t Bochum, Universit\"{a}tsstra\ss{}e 150, 44801 Bochum, Germany}

\date{\today}

\begin{abstract}

We report on the experimental observation of selective delamination of semi-transparent materials on the example of yttria-stabilized zirconia ceramics upon femtosecond laser processing of its surface with low numerical aperture lens. The delamination of a ceramic layer of dozens of micrometers takes place as a by-side effect of surface processing and is observed above the surface ablation threshold. The onset of delamination (delamination threshold) depends on the degree of overlap of the irradiation spots from consecutive laser pulses upon beam scanning over material surface. Analysis of the delaminated layer indicates that the material undergoes melting on its both surfaces. The mechanism of delamination is identified as a complex interplay between the optical response of laser-generated free-electron plasma and nonlinear effects upon laser beam propagation in semi-transparent ceramics. The discovered effect enables controllable laser microslicing of brittle ceramic materials. 

\end{abstract}


\maketitle

\section{I. Introduction}

Ultrashort-pulse laser micromachining of materials is attracting growing interest due to the possibility of achieving much higher quality of the laser processed surfaces as compared to longer laser pulses \cite{Malinauskas.2016}. The enhanced processing quality is largely conditioned by the difference in the laser ablation mechanisms for ultrashort (at the range from femtoseconds to dozen of picoseconds) laser pulses as compared to longer pulses. The difference results from strong thermal and stress confinements inherent for ultrashort laser pulses \cite{Paltauf.2003}. Although generally the conditions of stress confinement can be achieved at different pulse durations \cite{Paltauf.2003}, at ultrashort laser pulses this effect is more distinct so that the ablation can occur in the form of mechanical fracture and ejection of a layer of the irradiated material (referred to as spallation) as a result of the development of tensile stresses \cite{Zhigilei.1999,Ivanov.2003}. 

Spallation of laser-irradiated materials from front target surface and from rear surface in the case of films/foils has been extensively studied both experimentally \cite{Fox.1973,Gilath.1988,Eliezer.1990,Tinten.1998,Tamura.2001,Koch.2005,Savolainen.2011,Puerto.2012} and theoretically \cite{Eliezer.1990,Zhigilei.1999,Ivanov.2003,Zhakhovskii.2008}. In all experiments on the rear surface spallation of metals, the irradiation spot size was considerably larger than the  metal film thickness \cite{Fox.1973,Gilath.1988,Eliezer.1990,Tamura.2001}. It has been proven that the rear-side spallation effect is conditioned by the reflection of the laser-induced shock wave from the rear surface with the formation of a region with a high strain rate sufficient for creation of voids/cracks. For the case of front surface spallation, which can also manifest itself as swelling, the spalled layers were found to be of a submicrometer thickness with the size dependent on the materials properties \cite{Koch.2005,Savolainen.2011,Puerto.2012}. 

In this paper, we report on front side delamination of a layer from yttria-stabilized zirconia (YSZ) foil upon femtosecond laser processing of its surface with a low numerical aperture lens. Here we call this effect 'delamination' to underline the difference from the spallation mechanism mentioned above. In our case, delamination takes place from the front side of the target irradiated with multiple laser pulses at fluences $F$ exceeding the surface ablation threshold. The delaminated layer thickness is from ten to several dozens of micrometers that depends on the irradiation conditions. It must be underlined that the irradiation spot size on the material surface is much smaller as compared to the thickness of the YSZ samples. We discuss the physical mechanisms of this effect and demonstrate that delamination happens due to a complex interplay of the two major phenomena, self-focusing of the laser beam transmitted toward the material bulk and its defocusing by the free-electron plasma generated in the surface layer of the sample.

It is noteworthy that recently Kim et al. \cite{Kim} have reported on using femtosecond laser pulses for slicing 4H-SiC wafers by tight beam focusing (NA = 0.8) to a desired depth inside the sample. By using this method, exfoliation of $\sim$260 $\mu$m 4H-SiC layer was achieved with smaller roughness and material losses as compared to conventional slicing techniques. The laser irradiation conditions used in the present study differ considerably from those of work \cite{Kim}: the beam was focused on the sample surface with a low numerical aperture lens. However, due to highly non-linear properties of YSZ ceramic, the observed delamination effect that is analyzed below can have some analogy with the that reported in \cite{Kim}.

\section{II. Experimental Setup}

Experiments were performed with the diode-pumped fiber laser {\it Tangerine} produced by {\it Amplitude Systems}, emitting in TEM$_{00}$-mode at a wavelength of 1030\,nm and a pulse duration of 290 fs (full width at half maximum, FWHM). Repetition rate is scalable up to 2\,MHz. The laser beam quality factor M$^2$ is about 1.15 and the beam diameter $D_{\textrm{FWHM}}$ is about 1.25\,mm at the laser output. The downstream beam expander  by {\it Thorlabs} expands the beam diameter by the factor of 3, to 3.75\,mm. The beam is directed by a set of mirrors to the Galvo-scanner {\it SCANcube 7} by {\it SCANLAB}. At the scanner output, an F-theta lens with focal length of $f = 63\,$mm (numerical aperture 0.06) by {\it SCANLAB} focuses the beam on the sample surface located on the XYZ stage with z-axis parallel to the laser beam. Processed areas were either $0.5\times0.5$ mm$^2$ or $5\times 5$ mm$^2$. The beam waist in the focal plane is $w_0 \approx 7.5\,$\textmu m and the Rayleigh-length $z_R\approx$ 149 \textmu m is close to the thickness of the irradiated sample.


The yttria stabilized zirconia (or 8YSZ) used in these studies is zirconium dioxide with 8\% of yttrium oxide molar percentage, which is added to stabilize the cubic lattice. We notice however that recent publications indicate that a complete stabilization at room temperature is not achieved and there are inclusions of tetragonal phase, so called t" \cite{Butz.2011}.
The melting point of 8YSZ is about 2700$^\circ$ C. Further material data is given in Table \ref{material}. The  samples were provided by {\it Forschungszentrum J{\"u}lich GmbH, IEK1} purchased from {\it KERAFOL}. Dimensions of the samples are 25$\times$25$\times$0.2 mm$^3$.

\begin{table}
	\centering
	\caption{Physical properties of 8YSZ. }		
	\begin{tabular} { |l|c|c|l|}
		
		\hline
		Characteristics 	& Formula symbol			& Value 		& Unit \\ 
		\hline
		Crystal lattice \cite{Selcuk.1997}				& - 			& cubic	& - \\
		Band gap \cite{Goetsch.2016}						& $E_g$			&5.3	&eV \\
		Density\footnote{ Manufacturer specification}	& $\rho$		& 5950	& kg~m$^{-3}$ \\	
		Absorbtion depth at 1030nm\footnote{Obtained for virgin samples by measuring transmission and reflectance using \textit{UV-3600plus} by SHIMADZU} 			& $l_{a}$ 		& 53 	& $\mu$m \\
        		Reflection coefficient at 1030nm$^b$ 			& $R$ 		& 0.52 	&  \\
		Heat capacity \cite{Vaen.2004} ($20^\circ C$)	& $c_{p}$		& 500 	& J~kg$^{-1}$K$^{-1}$ \\
		Heat capacity \cite{Vaen.2004} ($1200^\circ C$)	& $c_{p}$		& 670 	& J~kg$^{-1}$K$^{-1}$ \\
		Thermal conductivity \cite{Vaen.2004}			& $\lambda_{th}$& 2.2	& W~m$^{-1}$K$^{-1}$ \\
        Thermal diffusivity ($20^\circ C$)				& $\kappa$		& $7.4\cdot 10^{-7}$	& m$^{2}~$s$^{-1}$\\
		\hline
	\end{tabular}
	
	\label{material}
\end{table}

\section{III. Results and Discussion}

\subsection{A. General features of 8YSZ processing}

Depending on the processing parameters (laser fluence and overlap of the irradiation spots upon scanning) different modes of sample modification/ablation are observed. We define the ablation threshold as the highest fluence at which no visible modification is seen in white-light interferometry (WLI). Additionally, we introduce a threshold fluence for the onset of delamination, $F_{th}^\textup{delam}$. This threshold is determined as the lowest investigated fluence, at which a layer of delaminated material can be found on the top of the laser processed area, though this layer can cover the processed area only partially. These two thresholds are slightly varying from one to another set of the experiments that can be conditioned by the initial sample defects. They depend on the scanning speed and the pulse repetition rate. As will be shown below, varying the scanning speed can compensate, to a certain extent, the change in the repetition rate to ensure a similar overlap (OL). However, the delamination threshold and the delaminated layer thicknesses can be slightly different for the same overlaps at different repetition rates that can be referred to the heat accumulation effect. Below we give the thresholds as the ranges of fluence in which either ablation or delamination (ablation or delamination threshold respectively) start to be observed in all our experiments.

\begin{figure} [h]
	\includegraphics[width=16cm]{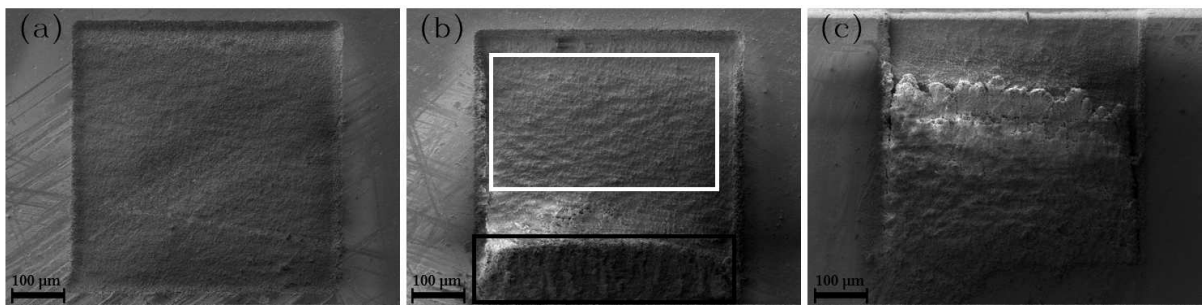}
    \caption{SEM images of the laser processed areas. (a) OL = 22\% and $F=20.1$ J~cm$^{-2}$. No delamination features are observed. (b) OL = 75\% and $F=24.9$ J~cm$^{-2}$.The most part is clean from any particulates (outlined by white frame) while the delaminated layer can be seen in the black-framed region. (c) OL = 83\% and $F=8.8$ J~cm$^{-2}$. Delaminated layer covers a substantial part of the processed surface (bottom part of the image) and its boundary in the form of flakes is clearly seen at the top part of the image.}
	\label{onset}
\end{figure}

Figure \ref{onset} demonstrates typical scanning electron microscope (SEM) images of laser-processed surfaces when delamination is either not observed or partially seen. Between the two above-defined thresholds, the processed area does not contain microparticulates and is relatively smooth as demonstrated in Fig.~\ref{onset}(a). At fluences just above the delamination threshold for a particular overlap, the delaminated layer covers the processed surface only partially as shown in Fig.~\ref{onset}(b)-(c). It is always attached to the side of the processed area where laser scanning was started. That can be an indication of an accumulation effect, which leads to destroying the delaminated layer upon continued scanning. The roughness of the unprocessed YSZ surfaces is 0.15 $\mu$m. The roughness of the processed areas in Fig.~\ref{onset}(a) and Fig.~\ref{onset}(b) (in white-framed area) was 0.5 $\mu$m and 2 $\mu$m respectively.

\begin{figure} 
	\includegraphics[width=12cm]{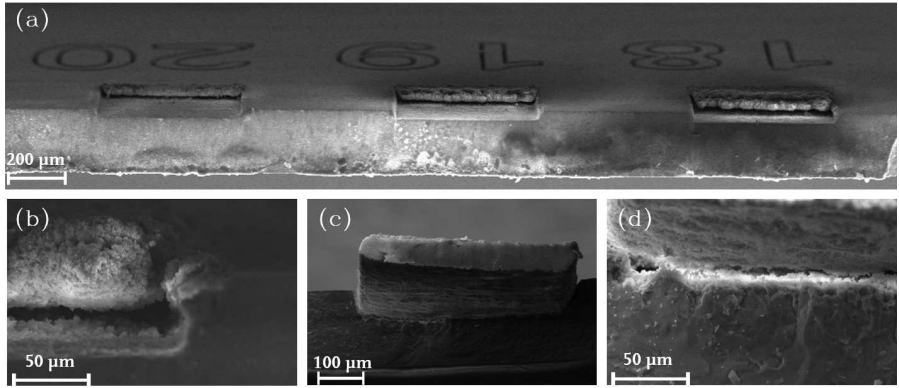}
	\caption{SEM images of laserprocessed areas with different fluences. Overlap (83.1\%) and repetition rate (100kHz) were constant for all cases shown. (a) Cross sectional view of the laser-processed sample (30$^\circ$-tilted view). Laser fluences are 7.2 J~cm$^{-2}$, 7.9 J~cm$^{-2}$ and 8.6 J~cm$^{-2}$ (marked as 20, 19 and 18 respectively). (b) Magnified view of the edge of the processed area (a) 18 in its contact with virgin sample area. Melting and ablation features can be recognized. (c) 30$^\circ$-tilted view of the delaminated layer at a laser fluence of 17.2 J~cm$^{-2}$. (d) Magnified view of (c), showing the contact between the delaminated layer and the underlying sample.}
	\label{overview}
\end{figure}

With further increasing fluence or overlap, the area becomes fully covered by the delaminated layer and the thickness of the layer increases as clearly seen in Fig.~\ref{overview}(a) (areas marked by numbers 20, 19, and 18). The figure presents the cross sectional image of the processed with the delaminated layers. For obtaining this image, the sample was cut across the processed areas. It should be noted that, in these cases of delamination, all three surfaces are modified with clear signs of melting, on both top and bottom of the delaminated layer as well as the surface from which delamination occurred. Also it can be noticed that the delaminated layer somewhat raises up from the initial sample surface. Figure \ref{overview}(b) presents the magnified view of the contact between the delaminated layer and the underlying sample as well as the edge where processed and virgin sample areas are contacting. The SEM image in Fig.~\ref{overview}(c) shows a free standing delaminated layer obtained due to a particular breakout at the cutting edge. The perspective is 30$^\circ$ tilted from the upright position. Fig.~\ref{overview}(d) supports that the bottom surface of this delaminated layer is modified.

\begin{figure} [h]
	\includegraphics[width=12cm]{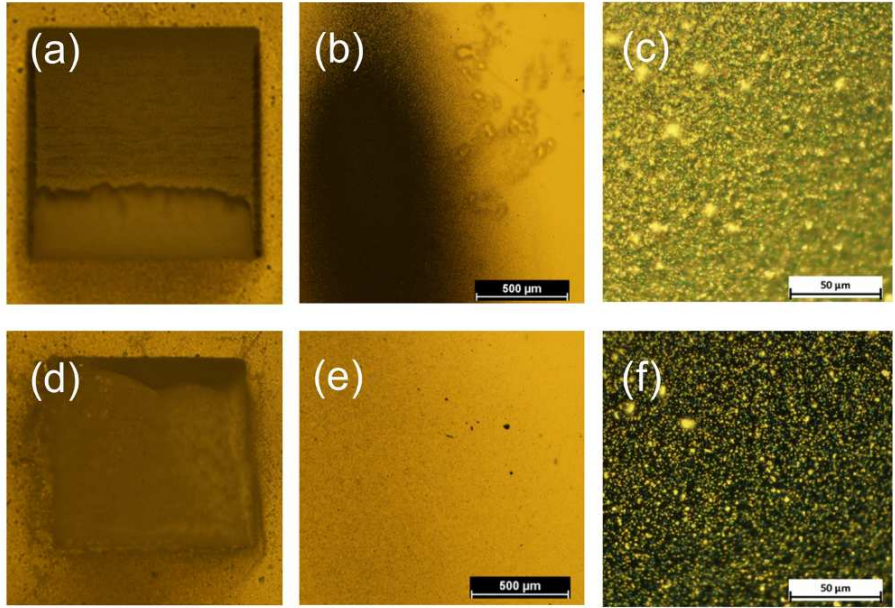}
	\caption{(a)-(c) Images of laser processed area at fluence of 1.6 J~cm$^{-2}$ obtained by optical microscopy. (a) partly destroyed delaminated layer, (b) corresponding glass substrates with the deposit of the ablation products in bright field and (c) in dark field. The delaminated layer is partially destroyed and its large fragments are clearly seen on the substrates. (d)-(f) The same for laser fluence of 5.7 J~cm$^{-2}$ when the whole delaminated layer remains attached to the sample. Only small rare particulates can be recognized on the substrate. OL = 92\% for all images.}
	\label{deposite_sm1}
\end{figure}

Here we hypothesize that, below the delamination threshold, the observed ablation of ceramics also proceeds via delamination but the delaminated layer is too thin. As a result, it is destroyed, most probably mechanically, via cracking and ejection from the processed sample. With increasing the thickness, the delamination layer withstands cracking and remains attached to the sample. This assumption can be verified by the inspection of the deposit of the ablation products on a collecting substrate. To do this, a microscope glass substrate was placed slightly aside of the laser beam, still assuring the laser plume deposition. The results are shown in Fig.~\ref{deposite_sm1} for the irradiation spot overlap of 92\%. The upper row presents images of the processed area (a) and glass substrate in bright (b) and dark (c) fields for laser fluence of 1.6 J~cm$^{-2}$. 
The delaminated layer is partially destroyed and consequently its large fragments are abundantly deposited on the substrates. At the same time, at laser fluence of 5.7 J~cm$^{-2}$ when the delaminated layer remains completely attached to the sample (d), only small rare particulates can be recognized on the glass substrate. 

Below we will address this assumption in more details and will provide the most probable physical mechanism of the observed delamination phenomenon.

\subsection{B. Ablation depth and delaminated layer thickness}

Based on the above assumption that the visible ablation naturally transfers to delamination with increasing laser fluence (note that the delaminated layer remaining on the sample after processing can be mechanically removed from the processed area), we introduce here a \textit{unified ablation depth} as the difference between the levels of virgin surface and the bottom of either ablated (without delamination signs) or delaminated area. The unified ablation depth was determined by white-light interferometry, using the \textit{Polytec TMS 1200} with Mirau-objectives. According to the device specification, the uncertainty in the step measurement varies between 0.18 $\mu$m and 0.1 $\mu$m depending on the step size. The reproducibility of the method was tested by measuring 15 different areas made on different YSZ samples with the same processing parameters. The standard deviation of the measurements was approximately 0.8 $\mu$m, which reveals certain deviation in the studied YSZ samples. For areas, which were not fully covered by the delaminated layer, it was measured in the regions free of delamination features. For the processed areas completely covered by the delamination layer, the corresponding cross sectional images were analyzed. 

\begin{table} 
	\centering
	\caption{Fluence and peak power ranges, starting from which delamination is observed. The data are presented for several overlaps and the two repetition rates, 200 and 100 kHz. The corresponding unified ablation depths as well as pulse-to-pulse and line-to-line shifts, $\Delta$x and $\Delta$y respectively, are also given.}
    
	\begin{tabular} {|c|c|c|c|c|c|c|}		
		\hline
		Overlap	&$\Delta$x	&$\Delta$y	& $F_{th}^\textup{delam}$ & Peak power	& Unified ablation	& Repetition rate \\ 
		in \% 	& in \textmu m	& in \textmu m	& in J~cm$^{-2}$  & in MW      & depth in \textmu m	& in kHz\\
		\hline
		92	&1			&1			& 2.2 - 2.6	   & 13 - 15 	& 11.2		& 200 \\
		83	&2			&2			& 4.3 - 4.9	   & 26 - 30	& 15.7		& 200   \\ 
		75	&3			&3			& 9.6 - 10.4   & 59 - 63	& 18.4		& 200    \\ 
		\hline
		75	&3			&3			& 12.7 - 13.5  & 78 - 82	& 22.8		& 100    \\
		67	&4			&4			& 22.9 - 23.9  & 140 - 146	& 23.1		& 100   \\		
		\hline
		
	\end{tabular}
	
	\label{treshold}
\end{table}

Table~\ref{treshold} outlines the tendencies of the unified ablation depth evolution as a function of overlap between irradiation spots. The ablation depths are given at the delamination thresholds indicated as the fluence ranges (see comment above) for two repetition rates, 100 and 200 kHz. The overlaps in x- and y-direction are always the same as indicated in the Table by sample shifts $\Delta$x and $\Delta$y between two subsequent pulses along the scanning line and between subsequent scanning lines respectively. All processed areas here were the same, 0.5$\times$0.5 mm$^2$. Decreasing overlap requires a higher fluence to observe the delaminated layer. Similarly, when decreasing the repetition rate with preserving the overlap, higher fluences have to be applied to obtain the delaminated layer attached to the processed surface. We note here that $F_{th}^\textup{delam}$ is typically several times higher than the ablation threshold. Thus, for 75\% overlap and 200\,kHz, the ablation threshold was found to be in the range of 2.2 - 2.6\,J~cm$^{-2}$, four times smaller than $F_{th}^\textup{delam}$ at this regime of processing, see Table~\ref{treshold}.

\begin{figure} [h]
	\includegraphics[width=8cm]{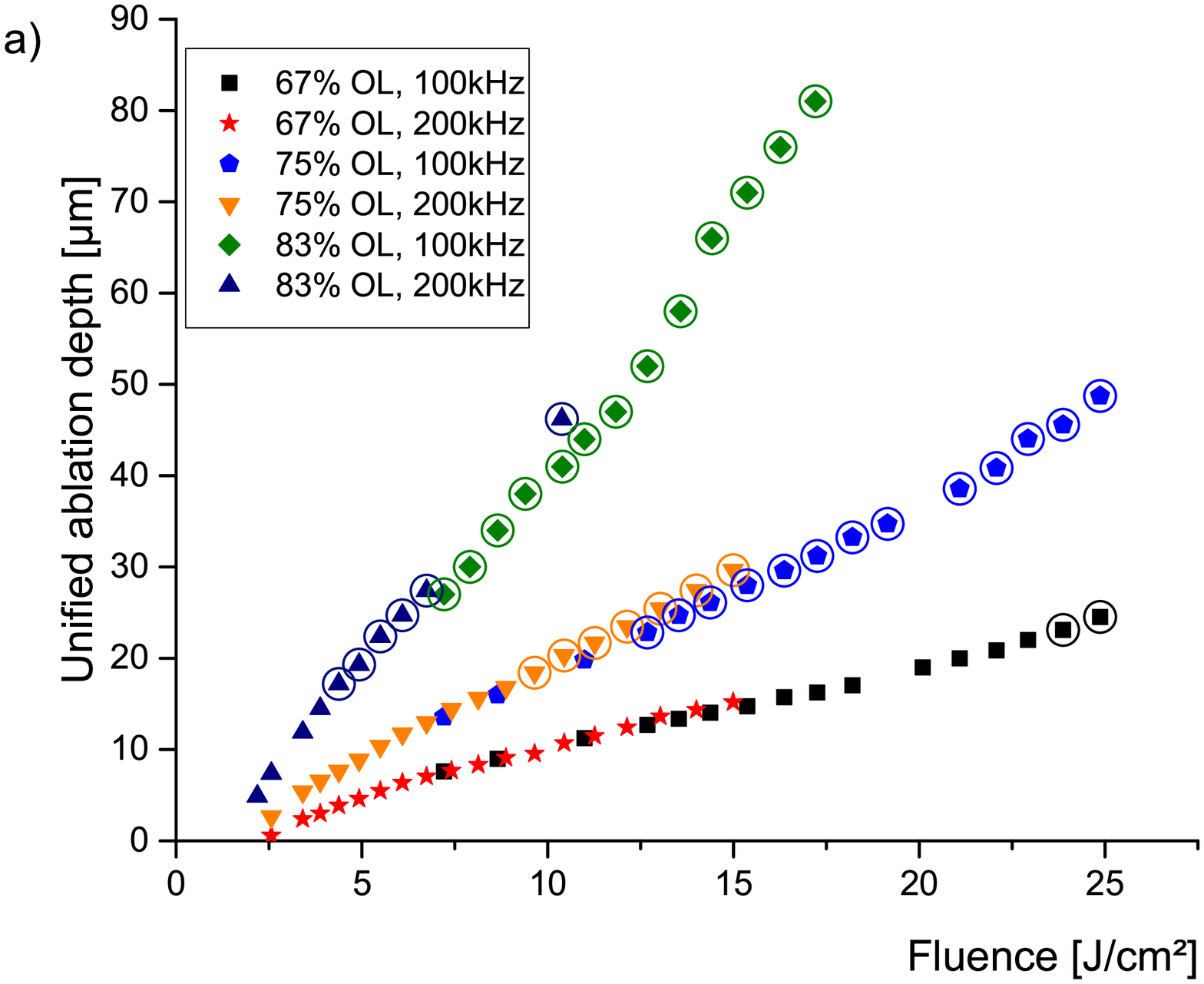} \includegraphics[width=8cm]{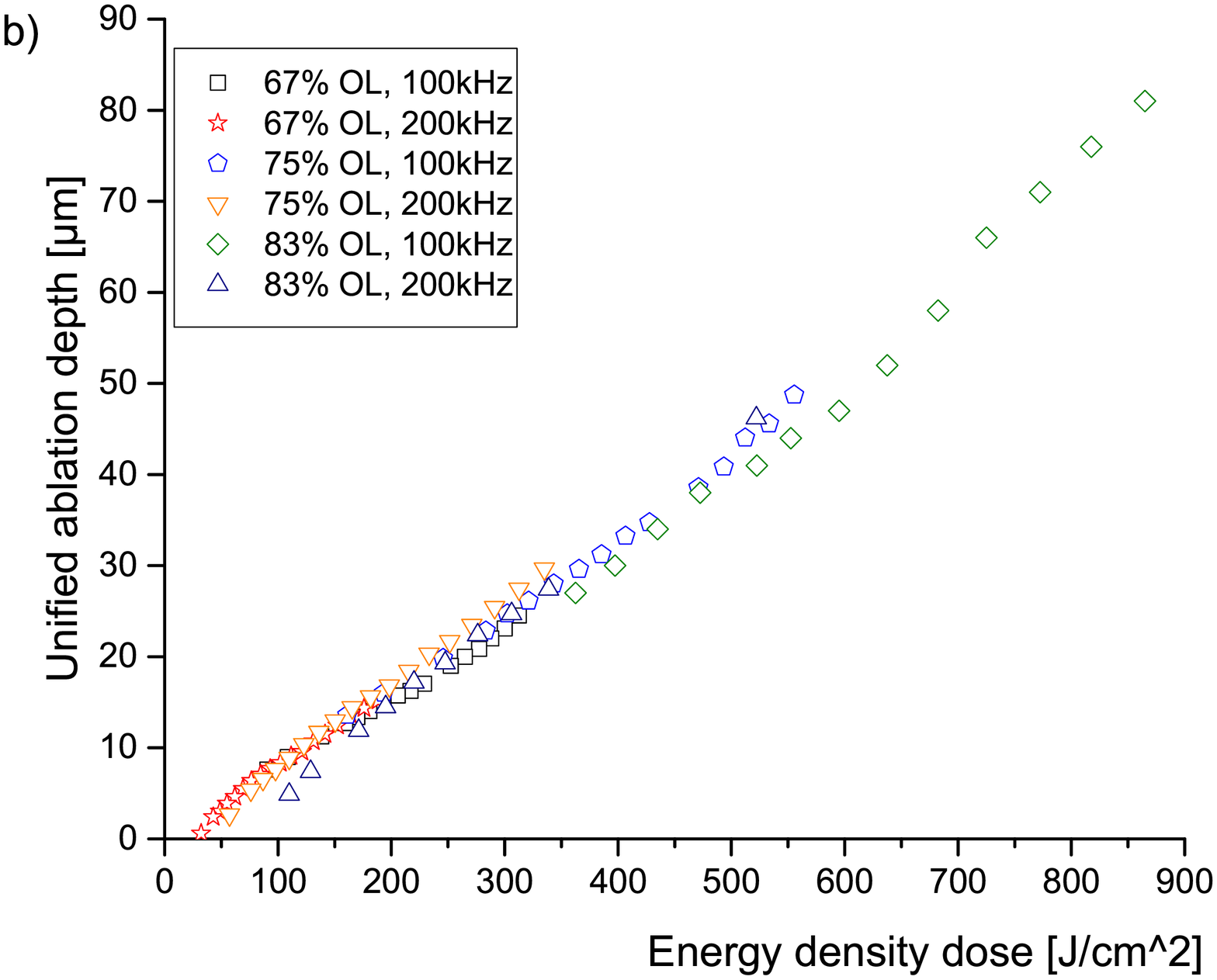}
	\caption{(a) Unified ablation depth as a function of laser pulse fluence for different overlaps and repetition rates. Circled datapoints correspond to fluences for which a delaminated layer could be observed. (b) Unified ablation depth as a function the energy density dose. No significant difference in the ablation depth could be observed for similar energy doses when varying parameter set of OL and fluence.}
	\label{thickness}
\end{figure}

Figure~\ref{thickness} presents the unified ablation depth (see definition above) as a function of (a) fluence and (b) energy density dose $\Theta$, which is defined as the product of single-pulse energy and the total number of pulses per area divided by the processed area size. All the results presented were obtained with two repetition rates, of 100\,kHz and 200\,kHz and three overlaps between pulses and lines, 67\%, 75\%, and 83\%. Symbols outlined in Fig.~\ref{thickness}(a) by circles refer to the processing regimes above the delamination thresholds. Based on Fig.~\ref{thickness}, several features of the ablation/delamination process can be highlighted:

\begin{itemize}
\item [-]The depth increases monotonously for increasing fluence (Fig.~\ref{thickness}(a)) and increasing energy density dose (Fig.~\ref{thickness}(b)). There are no visible peculiarities which would indicate a transition from ``pure" ablation to delamination.

\item [-] Increasing overlap at a constant single-pulse fluence considerably increases the unified ablation depth.

\item [-] Twice increasing the repetition rate has a tendency to slightly increase the unified ablation depth.

\item [-] The energy density dose defines the unified ablation depths as clearly seen from Fig.~\ref{thickness}(b).

\item [-] A similar unified ablation depth can be achieved by different parameter sets of overlap and fluence which correspond to the same energy density dose. As will be shown in Section III.D, the fraction of the delaminated layer, which remains attached to the processed sample, depends on the overlap and the number of scans.

\item [-] Interestingly, in all cases the delaminated layer survives on the processed area when the ablation reaches approximately 20 $\mu$m (somewhat smaller for higher overlap and slightly larger for smaller overlap).
\end{itemize}

The last feature indicates that the delaminated layer thinner than $\sim$20 \textmu m experiences cracking, fractures and is ejected from the processed surface as proven in Fig.~\ref{deposite_sm1} where large fragments of ceramics deposited on a collecting substrate are demonstrated for the processing regimes below $F_{th}^\textup{delam}$. Upon reaching a certain depth, the layer becomes able to withstand against cracking and remains attached to the sample. As a whole, the outlined features count in favor of delamination/fracturing of essentially mechanical nature whose mechanism will be addressed in Section III.F.

\subsection{C. Structure of the delaminated layer}

\begin{figure}
	\includegraphics[width=12cm]{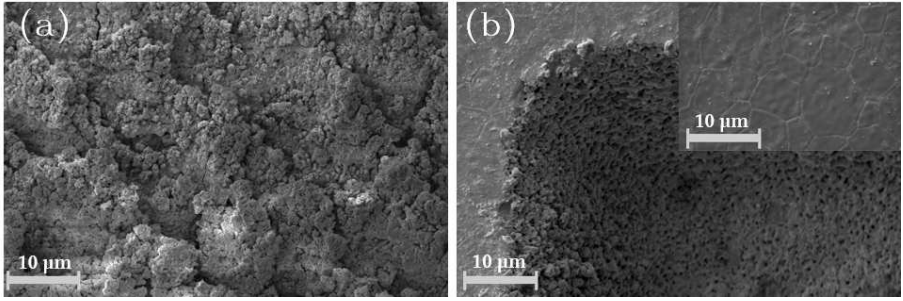}
    \caption{(a)Typical view of the ablation features on the top of the delaminated layer. (b) Morphology of the sample surface beneath the delaminated layer. The inset shows an unirradiated sample surface with observable grain boundaries. Images obtained by scanning electron microscopy. Overlap between the irradiation spots is 75\%; laser fluence is 17.2 J~cm$^{-2}$.}
	\label{ablation}
\end{figure}

Figure~\ref{ablation} presents the typical SEM images of the top surface of the delaminated layer and of the sample beneath it. It is apparent that ablation/modification of material occurs both on the top of the delaminated layer and at its contact with the rest sample. The inset shows an unirradiated area with the grain sizes significantly larger than the particles seen on the irradiated surfaces. Remarkable is that the signs of the initial grained structure can be recognized in the ablation relief, see Fig.~\ref{ablation}(a). The reason is a high concentration of defects in grain boundaries that provokes a preferred ablation at the boundary sites \cite{Ribeiro1997}. Hence, this image supports considerable ablation from the external surface of the delamination layer which must be mediated by the laser-induced breakdown and free-electron plasma formation at the sample surface \cite{Stuart.1996}. 

XRD (x-ray diffraction) measurements were carried out with the \textit{X\`{}Pert Pro} by \textit{PANalytical}, used in Bragg-Brentano geometry. XRD characterization of unprocessed and processed areas, the latter in the regimes with and without delamination, showed no differences indicating that the processed surfaces, including the delaminated layer, have the same polycrystalline structure as the unprocessed material.

\subsection{D. Repetition rate and heat accumulation}

\begin{figure}
	\includegraphics[width=12cm]{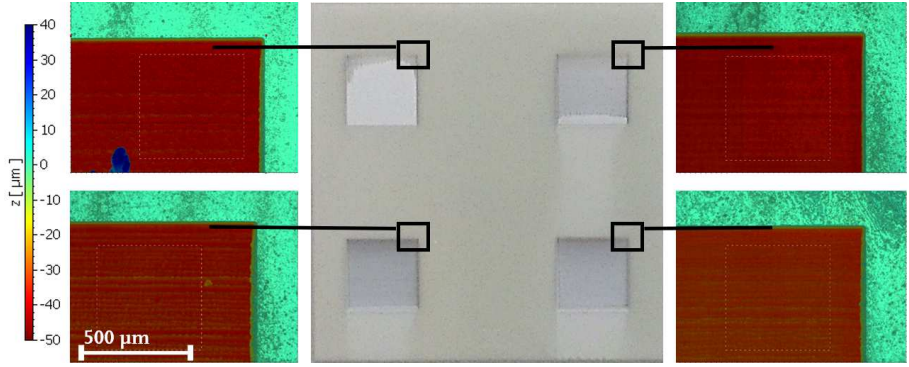}
	\caption{5$\times$5 mm processed areas irradiated with $F=11$ J~cm$^{-2}$ (photograph in the middle). Repetition rates were 100 kHz (upper-left, maximum depth of processing, $H$, is 40.3\,\textmu{}m), 50 kHz (upper-right, $H$ = 38.3\,\textmu{}m), 25\,kHz (lower-left, $H$ = 35.9\,\textmu{}m) and 10\,kHz (lower-right, $H$ = 34.8\,\textmu{}m). The decrease in the repetition rate was compensated by the scanning velocity to keep the same overlap of 83\%. The delaminated layers appear as lighter regions within the process areas. Note that processing was performed horizontally with gradual shifting the scanning lines from the bottom to the top in each processed area. Magnified views represent the WLI images. Scale and color map are applicable to all four WLI images.}
	\label{rep}
\end{figure}

As already indicated in Table~\ref{treshold} and Fig.~\ref{thickness}, a lower repetition rate results in a higher $F_{th}^\textup{delam}$. This can also be seen in Fig.~\ref{rep}. For all four processed areas in the middle gray-scale image, pulse fluence (11 J~cm$^{-2}$) and overlap (83\%) were the same and only repetition rate was changed in this series of laser processing from the upper left image to the lower right one from 100 to 50, 25 and 10 kHz, respectively. The ablation depths for these processed areas are 40.3, 38.3, 35.9 and 34.8 \textmu m respectively. The noticeable decrease in  the ablation depth with decreasing the repetition rate can be attributed to the heat accumulation effect. The heat accumulation at the surface layer of the sample should lead to the thermal expansion of the layer and correspondingly to a somewhat lower refractive index in the heat affected zone. This should consequently result in some defocusing of the laser beam part penetrating toward the sample bulk. According to the mechanism of delamination proposed below in Section III.F, beam refocusing deep in the sample produces the delamination cut. The surface layer expanded due to the heat accumulation effect can shift the cut deeper to the bulk at the higher repetition rates. Although heat accumulation can also cause thermal lensing, the absorbed heat is mostly confined in the delaminated layer and its gradient across the beam radius can be insufficient to counterbalance the thermal expansion effect. We suppose here that the ejection of the material due to self-focusing cut in its depth does not occur at each laser pulse but happens periodically as a result of stress accumulation from several pulses, see Section III.F.
  
The delaminated layer is clearly seen as a lighter region within the processed square-shaped area in the images. It is almost completely preserved on the sample without destruction at the highest repetition rate of $f=100\,$kHz while for lower repetition rates the remaining delaminated layer considerably decreases in size (Fig.~\ref{rep}). It is worth mentioning that the light shadows beneath the bottom parts of the processed areas for 50, 25 and 10 kHz originate from redeposition of particulates from the ablation plume while at 100 kHz, when the delaminated layer is almost completely preserved on the processed area, no visible signs of particulate redeposition is observed. The most plausible explanation is seen again in the heat accumulation effect emerging at higher repetition rates. Indeed it is known that ceramics usually become less brittle at enhanced temperatures \cite{Gogotsi.1978}. Hence, the higher the repetition rate, the higher is the temperature in the delaminated layer and the longer this layer can withstand to fracturing upon laser scanning. Note that the delamination of the irradiated layer from the sample confines the absorbed energy within the layer, thus enhancing the heat accumulation effect. 

The important role of heat accumulation has also been confirmed in the following series of the experiments. In this series, the number of pulses per each processed area of 0.5$\times$0.5 mm$^2$ and the single pulse fluence (and hence, the energy density dose) were kept constant. For the first two processed areas, all pulses were applied in one scanning run with high overlaps between the irradiation spots, 83\% (0.97 s scanning time) and 75\% (0.52 s scanning time), which correspond to 62500 and 27777 pulses per area respectively. Two other areas were processed in four scanning runs but with smaller overlaps of the irradiation spots within each scan (67\%, 0.33 s time per scan and 51\%, 0.18 s time per scan). As a result, the total numbers of pulses per area were the same as for the first two areas, 62500 and 27777 respectively. However, in the case of fourfold scanning with smaller OL, the energy density dose was four times smaller in each scan as compared to single scan with high overlap. Additionally to lower heating in each scan, the heat accumulated during one scan has a time to partially dissipate by the time of the next scan. As a result, fourfold scanning provides colder conditions of material processing.

\begin{figure}
	\includegraphics[width=12cm]{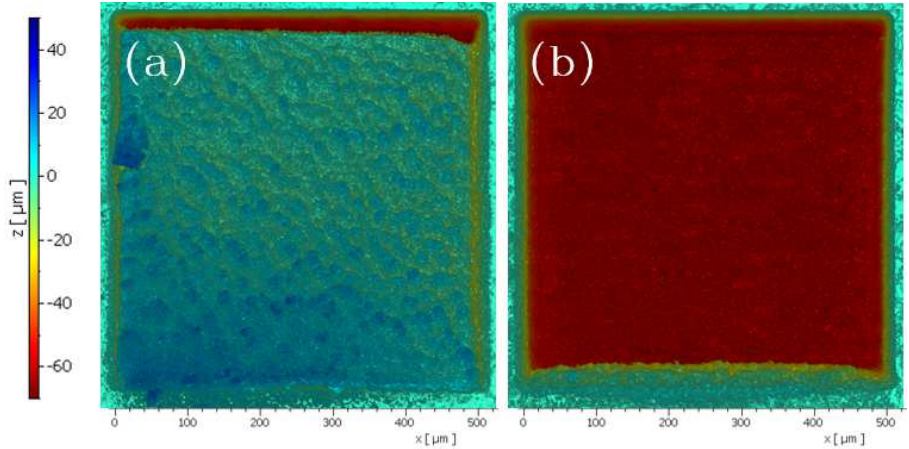}
	\caption{The areas of 0.5$\times$0.5 mm$^2$ processed with the same number of pulses, 62500. (a) Overlap 83\%, single scanning run. (b) Overlap 67\%, four scanning runs. In both cases, applied fluence was $F=15.4$ J~cm$^{-2}$, the energy density dose $\Theta=760$ J~cm$^{-2}$. Laser processing was started from the bottom edge of the image with scanning lines along $x$ direction. Images were obtained by WLI.}
	\label{62500}
\end{figure}

Figure \ref{62500} shows the WLI images of the processed areas of 0.5$\times$0.5 mm$^2$ with 62500 pulses. The applied single pulse fluence was 15.4 J~cm$^{-2}$ and the energy density doze was 760 J~cm$^{-2}$ for the processed area, resulting in the ablation depth of $\sim$61.4  $\mu$m in both single and fourfold scanning. However, at the single run, almost the whole delaminated layer has survived on the sample and only a very small region in the upper part of the processed area was evidently destroyed and ejected from the sample as seen in Fig.~\ref{62500}(a) (note that laser processing was started at the bottom edge of the image). On the contrary, at fourfold scanning with smaller overlap and time delays between subsequent scanning runs that should ensure better heat dissipation, only a small part of the delaminated layer have survived on the sample (Fig.~\ref{62500}(b)). These experiments demonstrate that heat accumulation  plays an important role in preventing destruction of the delaminated layer.

\begin{figure}
	\includegraphics[width=12cm]{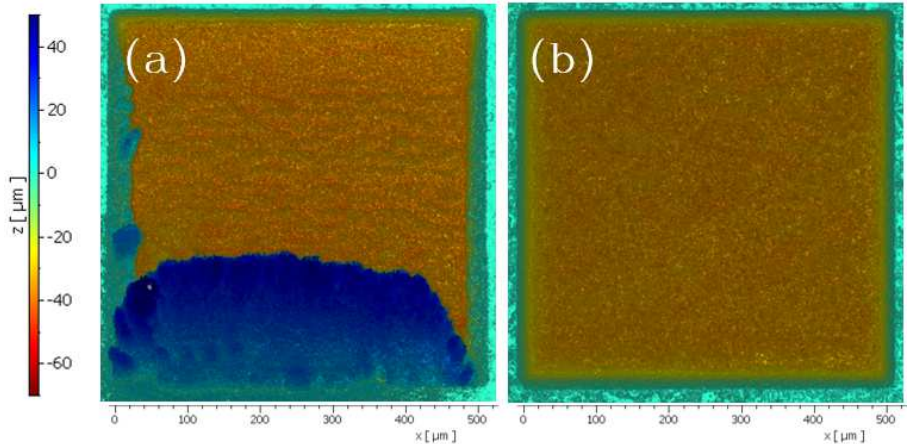}
	\caption{Same as in Fig.~\ref{62500} for 27777 pulses per 0.5$\times$0.5 mm$^2$ area. (a) Overlap 75\%, single scanning run; (b) Overlap 51\%, four scanning runs. Applied fluence was $F=20.1$ J~cm$^{-2}$, the energy density dose $\Theta=440$ J~cm$^{-2}$. Laser processing was started from the bottom edge of the image with scanning lines along x direction. Images are obtained by WLI.}
	\label{27777}
\end{figure}

Figure~\ref{27777} shows the WLI images of the processed areas of 0.5$\times$0.5 mm$^2$ with 27777 pulses at laser fluence of one pulse of 20.1 J~cm$^{-2}$ resulting in an energy density dose $\Theta=440$ J~cm$^{-2}$ for the processed area. On the surface processed by single scanning run with overlap of 75\%, the delaminated layer is partially preserved, being attached to the edge from which scanning was started (Fig.~\ref{27777}(a)). The ablation depth is 36.1 \textmu m in this case. In the area processed four times with 51\% overlap, the ablation depth is somewhat smaller than at 75\% overlap, 34.6 \textmu m (but comparable in regard to the standard deviation of 0.8 \textmu m of measurements, see section III B.), while no signs of the delamination layer are visible (Fig.~\ref{27777}(b)). Also it can be noticed that the delamination layer is raising up from the sample surface by about 40 \textmu m. It looks like a flake, being attached to the sample at the starting edge of processing and lifted off at the rest area, that is plausible due to pushing forces upon the delamination cut/crack formation (Fig.~\ref{27777}(a)).

Figures \ref{62500} and \ref{27777} clearly indicate the dominating role of the  energy density dose in regard to the ablation depth and the heat accumulation effect in regard to the delaminated layer stability. Although the laser fluence is higher for smaller number of pulses applied to the same area, the ablation depth and, hence, the delamination layer thickness is considerably smaller. To explain this and other features of the delamination effect, below we consider the processes taking place upon laser beam coupling to bandgap materials and discuss possible mechanisms and scenarios of the delamination effect.

\subsection{F. Possible mechanism of observed delamination: Counterbalancing between self-focusing and electron plasma anti-waveguiding}

\begin{figure}
	\includegraphics[width=15cm]{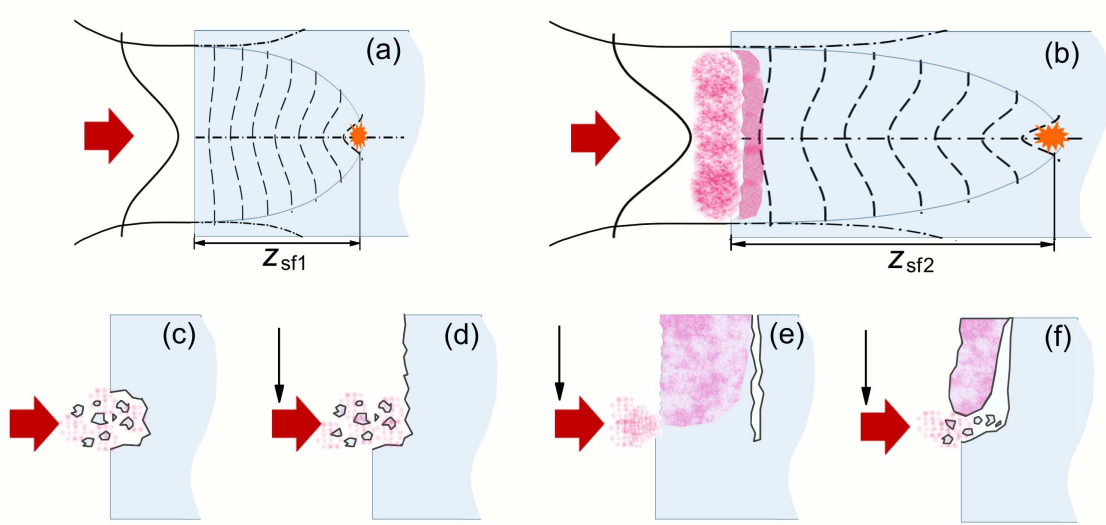}
\caption{(a) Illustration of laser beam self-focusing in a transparent non-linear solid with a high ionization threshold. The beam is focused on the sample surface. Instead of diverging after the geometrical focus (dash-dotted lines), the beam experiences self-focusing governed by the Kerr effect until its collapsing at the distance $z_{\text{sf}1}$ which culminates with generation of free-electron plasma. The scheme has been adapted from \cite{Chekalin.2013}. (b) For semi-transparent materials like ceramics considered in this paper, free-electron plasma is already generated at the surface layer (pink surface region) that can lead to melting and ablation of the surface layer. Free electron population counteracts to the Kerr effect by adding a negative contribution to the refractive index, see Eq. (\ref{n change}). This ``anti-waveguiding'' effect \cite{Chekalin.2013} is stronger for higher energy of the beam. As a result, self-focusing is delayed in space and the self-focusing distance $z_{\text{sf}2}$ is increasing with laser beam power. (c)--(f) Schematics of the ablation/delamination mechanism. At relatively low laser power (but above self-focusing threshold), the beam collapse happens close to the surface (as in (a)), resulting in fracturing the region between the collapse spot and  the surface (c). In such regimes with scanning (scanning direction is shown by black arrow), mechanical fracturing with ejection of ceramic fragments is the main mechanism of ablation (d). At high laser power, generation of a dense electron plasma in a thin surface layer leads to partial reflecting of laser light and in the ``anti-waveguiding'' effect \cite{Chekalin.2013}. As a result, the laser beam fraction, which is transmitted through the electron plasma layer, collapses deep in the target and provides the material melting, ablation and fracturing inside the bulk that is seen as the layer delamination (e). At intermediate beam powers, the layer delamination can transform to layer fracturing upon scanning (f) as seen in Figs.~\ref{rep},~\ref{62500} and~\ref{27777}.}
	\label{self-focusing}
\end{figure}

Ceramic delamination can be explained by laser beam self-focusing upon propagation in the non-linear optical medium. In non-linear media, the refractive index $n$ depends not only on the frequency of electromagnetic field but also on the local field intensity of the laser beam $I(r,z,t)$ as $n = n_0 + n_2I(r,z,t)$ where $n_0$ and $n_2$ are the linear and non-linear (Kerr) refractive indexes and $r$ and $z$ are respectively radial and axial coordinates. For transparent crystals and glasses, the value of $n_2$ is typically positive and in the range of $10^{-16}-10^{-14}$~cm$^2$~W$^{-1}$ \cite{Weber1978}. The wave front of powerful laser beams with the intensity increasing toward the axis (e.g., Gaussian as in our case) is distorted during beam propagation in such non-linear medium, as schematically shown in Fig.~\ref{self-focusing}(a), due to decreasing phase velocity in higher refractive index regions \cite{Chekalin.2013}. As a result, initially parallel optical rays are converging toward the beam axis, culminating in catastrophic collapse at a distance $z_{\text{sf}1}$ after the entering of the laser beam into the medium. The critical laser power for self-focusing, which is derived from the balance between the angles of self-focusing $\theta_{\text{sf}}$ and beam diffraction $\theta_{\text{df}}$, $\theta_{\text{sf}} = \theta_{\text{df}}$, can be evaluated as $P_{cr} \approx 3.72 \lambda_0^2/(8 \pi n_0n_2)$ where $\lambda_0$ is the laser wavelength \cite{Marburger.1975,Couairon.2007}.     

For yttria stabilized zirconia at laser wavelength of 1030 nm, $n_0$ = 2.1236 and $n_2$ = 1.184$\times$10$^{-15}$~cm$^2$~W$^{-1}$. Using these data, the critical power for self-focusing can be evaluated as $\sim$0.63 MW, which is more than the order of magnitude smaller than the smallest threshold value of beam power for delamination, see Table \ref{treshold}. For transparent (low-absorbing) media, the propagation depth of the beam till its collapse can be estimated by the empirical expression \cite{Marburger.1975,Couairon.2007}
\begin{equation}
L_c = \frac{0.367z_R}{\sqrt{[(P_{in}/P_{cr})^{0.5}-0.852]^2-0.0219}}.
\label{self-focusing depth}
\end{equation}
Here $P_{in}$ is the power of incident laser beam. 
We suppose that in our experiments the layer is delaminated at the depth of self-focusing $L_c$. Assuming as the first approximation that material absorption is insignificant before beam collapsing, one can evaluate $L_c \approx$ 11 \textmu m at a fluence of 2.5 J~cm$^{-2}$. Interestingly, this pair of values coincides with the threshold of delamination at high overlap (92\%, see Table \ref{treshold}). We can presume that, at such low laser fluences, only small fraction of light is absorbed before beam collapsing and, hence, the depth of beam collapse is reasonably described by Eq. (\ref{self-focusing depth}). Indeed, for the intrinsic (linear) absorption depth of 53 \textmu m of YSZ ceramics (assuming absence of non-linear absorption), less than 20\% of the beam energy is absorbed at the distance of 11 \textmu m and the Kerr focus is shifted insignificantly. 

However, as follows from Eq. (\ref{self-focusing depth}), the self-focusing distance has to move closer to the sample surface with increasing beam power. Thus, for the fluence of 7 J~cm$^{-2}$ $L_c \approx$ 7.4 \textmu m that is more than 3 times smaller as compared to the unified ablation depth for the overlap of 83\% (Fig.~\ref{thickness}, 200 kHz repetition rate). Below we show that there is no contradiction as, for semi-transparent materials irradiated with loose beam focusing on the surface and at high repetition rates, other effects can contribute to the position of the self-focus. 

The transient and permanent changes of refractive index in laser irradiated materials can be caused by several factors which include the Kerr effect ($\Delta n_{\text{Kerr}}$), generation of conduction-band electrons ($\Delta n_{\text{CB}}$), heat accumulation ($\Delta n_{\text{th}}$), accumulation of defects ($\Delta n_{\text{def}}$), density change in the heat affected zone ($\Delta n_{\rho}$), and local stress ($\Delta n_{\text{P}}$) \cite{Waxler.1973,Sakakura.2011,Bulgakova.2013}:
\begin{equation}
\Delta n = \Delta n_{\text{Kerr}} + \Delta n_{\text{CB}} + \Delta n_{\text{th}} + \Delta n_{\text{def}} + \Delta n_{\rho} + \Delta n_{\text{P}}.
\label{n change}
\end{equation}
The contribution of the Kerr effect is positive, resulting in narrowing and, finally, collapsing the laser beam (Fig.~\ref{self-focusing}(a)). Upon beam collapsing, a high local intensity is achieved which is enough to create free electrons. Free electron population is produced via photo-ionization which can trigger collisional ionization starting from a certain level of free electrons \cite{Stuart.1996}: 
\begin{equation}
\frac{dN_e}{dt} = (\sigma_kI^k +\alpha_{\text{col}}N_eI)\frac{(N_{0}-N_e)}{N_0}.
\label{free electrons}
\end{equation}

Here $N_e$ is the density of free electrons, $N_0$ is the atomic density of unexcited material, $\sigma_k$ and $k$ are the coefficient and the order of multi-photon ionization respectively, and $\alpha_{\text{col}}$ is the coefficient of collisional ionization. The factor $(N_0-N_e)/N_0$ is added to account for available ionization centers at high ionization rates \cite{Burakov.2005}. It should be underlined that, at ultrashort laser pulses, the avalanche ionization can considerably contribute to generation of free electrons in bandgap materials. Thus, Lenzner at al. \cite{Lenzner.2000} have shown that in fused silica the avalanche process is developing already at laser pulses of 120 fs duration that leads to strong decreasing the processing quality compared to shorter laser pulses. Furthermore, numerical simulations \cite{RethfeldSiO2} have demonstrated that, for fused silica at 300 fs laser pulses, the avalanche process contributes noticeably to material ionization already starting from $\sim 2 \times 10^{13}$ W/cm$^2$ and Eq.~(\ref{free electrons}) is applicable at intensities $\gtrsim 4 \times 10^{13}$ W/cm$^2$ (see Fig. 3 in \cite{RethfeldSiO2}). Note that such intensities are typical for our experiments while smaller band gap of YSZ ceramics ($E_g$ = 5.3 eV against 9 eV for fused silica) should result even at lower intensities for free carrier generation and subsequent triggering the avalanche process.

As soon as free electron plasma is produced in the conduction band, it counteracts to the Kerr self-focusing ($\Delta n_{\text{CB}} <0$) and can even lead to the anti-waveguiding effect \cite{Chekalin.2013}. In our case, when the laser beam is loosely focused on the sample surface with generation of free electrons in the surface layer according to Eq. (\ref{free electrons}), the free electron plasma can considerably alter the beam coupling to the material via increasing reflectivity, light defocusing/scattering, and attenuating the beam along its propagation toward the material bulk after a partial reflection from the electron plasma at the surface layer. Within a surface layer where the laser beam is not yet strongly distorted by self-focusing and defocusing, attenuation of laser intensity can be roughly described in a one-dimensional form as 
\begin{equation}
\frac{dI}{dz} = -\alpha_{\text{in}}I - \sigma_kI^k\frac{(N_{0}-N_e)}{N_0}k\hbar\omega -\alpha_{\text{fe}}I.
\label{beam attenuation}
\end{equation}

Here $\alpha_{in} = 1/l_a$ is the intrinsic absorption coefficient and $\alpha_{\text{fe}}$ is the absorption coefficient of free electrons produced by the laser light. The optical response of the dynamically ionized dielectric target (both dynamic change of the reflection coefficient and the spatio-temporal behavior of $\alpha_{\text{fe}}$) can be calculated through the complex dielectric function $\epsilon(N_e)$ by involving the Drude theory \cite{Burakov.2005}. 

A dynamically evolving reflectivity of the beam from the sample surface and attenuation of its part penetrating toward the sample bulk have to strongly affect the position of the Kerr focus under the condition that the beam power remains in excess of $P_{cr}$ after the beam passes through the excited surface region. It can be stated that the Kerr focus is dynamic under such excitation conditions and its position depends on the fraction of the beam energy (power) that has passed through the free-electron-plasma ``shield" generated in the surface layer. Generally, the transient plasma ``mirror/attenuator" created in the surface layer of the sample should move the Kerr focus deeper to the material bulk as schematically shown in Fig.~\ref{self-focusing}(b). We recall that, at 7 J~cm$^{-2}$, $P_{in}/P_{cr} \approx 68$, yielding in $L_c \approx$ 7.4 $\mu$m as estimated by Eq. (\ref{self-focusing depth}). It is possible to roughly evaluate that, for moving the Kerr focus deeper to the sample, to $\sim27$ $\mu$m from the surface (Fig.~\ref{thickness}), the beam power must decrease by the factor of $\sim 8.3$ after partial reflection and attenuation by the free electron plasma at the surface layer (to achieve $P/P_{cr} \approx$ 8.1-8.2 after passing the plasma layer). Note that in such a case the laser fluence drops from 7 J~cm$^{-2}$ at the sample surface to a local level below 1 J~cm$^{-2}$, the latter is well smaller than the damage threshold of a wide bandgap dielectric (estimated direct bandgap of yttria-stabilized zirconia is around 5.2-5.8 eV \cite{Ostanin.2000}). Here under the damage any irreversible change of material is meant which is observed after laser irradiation such as visible signs of melting, material ablation, cracking, change of crystalline structure, compaction or appearance of porosity. Note that the minimal laser fluence starting from which the damage is observed (damage threshold) is scaling with the band gap of dielectric materials and for materials with $E_g >$ 5 eV it exceeds 1 J cm$^{-2}$ at pulse durations of 100 fs and longer \cite{Mero2005}. Upon self-focusing, the intensity of the attenuated laser beam can reach again the value sufficient for the free electron production, which in its turn will induce local material heating and stress generation, in analogy with \cite{Kim} where the laser beam was purposely focused inside 4H-SiC wafer with a high numerical aperture lens. 
Hence, the laser-induced free electron plasma created upon focusing the laser beam on the sample surface can reasonably explain the delamination effect and its depth found in this work.

Reliable numerical simulations of the experimental conditions presented here are not seen possible in view of extremely large computational resources required for such kind of problems (see, e.g., \cite{Bulgakova.2015}) and a number of unknown material parameters for description of free-electron generation. To estimate light reflection and absorption by the generated free electron plasma, we recall that, even for the materials with a larger band gap such as fused silica under similar surface-irradiation conditions, an overcritical free-electron density is produced within the laser fluence range used in the present experiments \cite{Mirza.2016}. According to simulations for fused silica (see Fig. 6 in \cite{Mirza.2016}) and taking into account intrinsic reflectivity of 8YSZ ceramics (Table \ref{material}), at laser fluence of $\sim$7 J~cm$^{-2}$ (peak fluence of $\sim$14 J~cm$^{-2}$) more than half of the laser energy is reflected from the surface. The rest of laser energy, which penetrates to the sample, is attenuated due to photo-ionization and absorption by free electrons during propagation toward the target. The attenuated energy density can exceed 3 J~cm$^{-2}$ at the distance of $\sim$4 $\mu$m. Note that, for this estimation, we assume that an average energy spent for free electron production and heating is in the range of 50-60 eV per electron which is consistent with simulations \cite{Mirza.2016} and experiments \cite{Geoffroy.2014}. Under such conditions of laser energy absorption, the laser beam is attenuated to the fluence below the damage threshold but still it has a power above $P_{cr}$ in respect of the self-focusing effect. Note that additionally the laser beam can be scattered (defocused) by the free electron plasma. In view of a lower band gap of yttria-stabilized zirconia as compared to fused silica, the generated free electron density can be even higher than considered in the above estimations that will result in a higher light absorption within the surface layer. Hence, the scenario presented in Fig.~\ref{self-focusing}(b) seems to be plausible and convincing: 

- With increasing beam energy, the laser-generated electron plasma in the surface layer of the sample strongly depletes the laser beam. The absorbed laser energy in this layer is enough to induce melting and even partial ablation at the sample surface. 

- The fraction of the beam, which passes through the free-electron region, is not sufficiently energetic to cause material damage. 

- However, as the power of the beam is still above $P_{cr}$, the beam experiences self-focusing with the Kerr focus well deeper to the sample as compared to what could be expected for the case of absent or weak absorption. As a result, new local region of high laser energy absorption appears in the Kerr focus, inducing new damage (melting, cracking, internal ablation). 

Regarding the ablation/delamination mechanisms, the following conclusions can be done based on the above considerations. At relatively low laser fluences (but above the self-focusing threshold in terms of pulse power), the free electron density generated in the surface layer of the sample as well as linear material absorption are insufficient to induce the surface damage. As a result, after partial reflection from the surface and absorption in the surface layer, the beam penetrates toward the bulk and collapses in the subsurface region. In the collapse region, due to formation of a highly localized free-electron population which transfers its energy to the lattice upon recombination at the time scale of few picoseconds, ceramics must melt and a very high stress is generated. It was shown that, in the beam focusing region deep inside fused silica bulk, the stress level is of the order of 70-80 MPa \cite{Bulgakova.2015}. In YSZ ceramics under similar focusing conditions, the maximum stress level is expected to be more than the order of magnitude higher. Indeed, the stress is proportional to Young's modulus (approximately 2.5-3 times higher for YSZ \cite{Adams.1997} as compared to fused silica) and the coefficient of thermal expansion ($\sim$10$^{-5}$ K$^{-1}$ for YSZ \cite{Hayashi.2005} against 0.55$\times$10$^{-6}$ K$^{-1}$ for fused silica) \cite{Kingery.1955}. As the tensile strength of YSZ ceramics is reported to be 745 MPa \cite{Noguchi.1989}, the expected stress has to considerably exceed the material strength, leading to mechanical damage around the collapse region. As estimated above, at relatively low beam powers and, hence,  at the absence or at low free-electron plasma shielding, the self-focus has to be formed close to the sample surface. The laser-induced stress, which exceeds the material strength, should cause fracturing of the material layer between the focus and the surface with ejection of particulates (Fig.~\ref{self-focusing}(c)). In such regimes, mechanical fracturing of the surface layer is the main mechanism of ablation upon laser processing (see Fig.~\ref{self-focusing}(d)) that was confirmed by the experiments with deposition of the ablation products (Fig.~\ref{deposite_sm1}). The smallest fluence, at which such material removal starts to be observed, is considered as the ablation threshold. Noticeable is that the ablation thresholds for different processing conditions (pulse repetition rates, irradiation spot overlap) differ insignificantly (Fig.~\ref{thickness}(a)). It is known that, at ultrashort laser irradiation of bandgap materials, the surface damage threshold drops dramatically with the number of pulses applied to the same surface area due to material-dependent incubation effects \cite{Rosenfeld1999}. An insignificant difference in the ablation thresholds at different overlaps upon laser scanning observed in our experiments supports that the ablation process is governed by beam self-focusing, which is weakly dependent on incubation effects in the surface layer of the sample. Nevertheless, it can be expected that, upon laser scanning, not every pulse leads to material fracturing but ejection of particulates happens periodically as a result of stress accumulation from several laser pulses. 

At high beam powers, a high-density free-electron plasma is formed at the very surface layer of the sample that results in shielding and ``anti-waveguiding" of the laser beam. The fraction of the beam, which penetrates through the shielding area, experiences collapsing at a large distance from the surface as discussed above. In such cases, the stress generated in the self-focus region, is not enough to induce fracturing of a relatively thick material layer between the self-focus and the surface. The high-temperature/high-pressure local zone inside the bulk evolves into a pore \cite{Bulgakova.2015}. Namely in such regimes, the delamination effect over the whole processed area is developing, which originates from the line of the adjacent or closely located pores (Fig.~\ref{self-focusing}(e)). At intermediate laser powers, the delaminated layer can be preserved on the sample till a certain level of the accumulated stress after which it is starts to fracture at further processing (Fig.~\ref{self-focusing}(f)). The area of the preserved delaminated layer depends on the depth of self-focusing and the overlapping degree, the latter determines the heat accumulation in the delaminated layer and, hence, its mechanical properties.  

It is necessary to admit that, due to multiplicity of factors influencing light propagation in materials (Eq. (\ref{n change})), the delamination effect uncovered in this work is a very complex phenomenon. Under multi-pulse irradiation of the same area with relatively high overlapping of the irradiation spots as in the present experiments, accumulation of heat and defect states (the latter are abundant in yttria-stabilized zirconia \cite{Foster.2002}) can change the self-focusing conditions via creation of the thermal and defect convex lenses. Indeed, both terms $\Delta n_{\text{th}}$ and $\Delta n_{\text{def}}$ in Eq. (\ref{n change}) are positive \cite{Waxler.1973,Sakakura.2011,Martin.1997} and can assist in the self-focusing effect. As for two last terms in Eq. (\ref{n change}), $\Delta n_{\rho}$ and $\Delta n_{\text{P}}$, their roles at multi-pulse irradiation are more complicated. Heat accumulation in the delaminated layer should lead to a decrease of material density within the layer ensuring $\Delta n_{\rho} < 0$. On the other hand, at each pulse within the light absorption region, material can be relocated with creation zones of higher and lower density as compared to the virgin one \cite{Bulgakova.2015}. It could be speculated that a pressure-induced compacted shell is created, which surrounds the pathway of the focused/self-focusing beam toward the focal region. This compacted shell, which is subjected also to the residual stress accumulation, should affect the propagation of the next laser pulse and, most plausibly, assists in the light guiding toward the plane of the beam collapse \cite{Chan.2003}. We note that the waveguiding effect of a transient lens created by a mutual action of thermal and defect-induced lenses as well as by the material compaction shell, must also be inherent for direct laser writing of waveguides in optical glasses.

\section{IV. CONCLUSION}

 In this paper we have analyzed ultrashort-pulse laser ablation of semi-transparent materials on the example of YSZ ceramics. Unlike transparent (e.g., fused silica) or strongly-absorbing (e.g., metals) materials, here the laser ablation process is strongly influenced by delamination of a relatively thick surface layer. The depth of the delamination (and hence the depth of the crater) depends on the interplay between Kerr self-focusing due to the positive nonlinear refractive index of the material and beam defocussing induced by free-electron plasma formation at the sample surface. As the free-electron plasma density is evolving during the laser pulse, the position where beam self-focusing may happen can also evolve in time. The actual position of the beam collapse is determined by the strength of a negative free-electon lens achieved during the pulse. When the incident pulse power increases, a denser free-electron `shield' is created on the sample surface that leads to a stronger anti-waveguiding effect and spatial delaying of the beam collapse. Our studies show that the unified ablation depth as a function of laser fluence (see Fig.~\ref{thickness}(a)) can be better fitted by the linear dependence than by a logarithmic one, inherent for thermal mechanisms of ablation \cite{Chichkov1996}. Together with the depth of observed delamination, which well exceeds the size of the laser irradiation spot, this supports the relevance of the proposed phenomenological model. This modeling representation reasonably explains the dependence of the ablation depth on the laser fluence and provides an adequate quantitative estimation for the crater depth at the threshold of delamination. 

At high laser fluences well exceeding the ablation threshold and strong overlaps between the irradiation spots upon processing, a paradoxical effect can be observed: no crater is left on the surface anymore but instead the processed area raises up from the virgin sample surface, see e.g. Fig.~\ref{62500}(a). It has been found that, in such cases, the delaminated layer is thick enough to withstand dynamic mechanical stresses and remains attached to the processed area as schematically shown in Fig.~\ref{self-focusing}(e).

Summarizing, in this study we have demonstrated laser-induced delamination of layers with the thickness of several tens of micrometers and the area of nearly 5$\times$5\,mm from the bulk YSZ ceramics. It has been shown that the delaminated layer thickness can be controlled by laser fluence and overlap of the irradiation spots upon laser scanning of samples. Consequently, the discovered effect opens up a new way for controllable laser microslicing of brittle ceramic materials, i.e. cutting two-dimensional high-aspect-ratio sheets parallel to the bulk surface. 

\section{ACKNOWLEDGMENT}

NMB acknowledges the European Regional Development Fund and the state budget of the Czech Republic (project BIATRI: No. CZ.$02.1.01/0.0/0.0/15\_003/0000445$, project HiLASE CoE: No. CZ$.02.1.01/0.0/0.0/15\_006/0000674$, programme NPU I: project No. LO1602). Delivery of starting materials by Forschungszentrum J\"{u}lich GmbH (K. Wilkner, M. Bram) is greatly acknowledged.

\end{document}